\documentstyle [11pt,bezier,epsfig]{article}
\begin{document}
\begin{titlepage}
\begin{center}
\begin{large}
EUROPEAN LABORATORY FOR PARTICLE PHYSICS (CERN)
\end{large}
\end{center}
\vglue 1.0cm
    \begin{flushright}
        ALEPH 98-048 \\
        CONF 98-022 \\
        \today \\
        \vspace{0.5cm}
        {\bf Abstract Registration Number 1038} \\
        {\bf Parallel session: 1,12} \\
        {\bf Plenary session: 1,14}
    \end{flushright}

\vspace{0.5cm}

\begin{center}
{\Huge{\bf PRELIMINARY }} \\
    \vspace{1cm}
{\LARGE{\bf  Determination of the\\ 
          LEP center-of-mass energy \\
           from Z$\gamma$ events } }
\end{center}

\vspace{1cm}

\begin{center}
  {\bf The ALEPH Collaboration } 
\end{center}
\vspace{2cm}

\begin{center}
{\bf   Abstract}
\end{center}
Radiative returns to the Z resonance (Z$\gamma$ events) provide a method to
determine the centre-of-mass energy of the high 
energy $e^+e^-$ colliders and already at LEP2. The method is
applied for the data collected with the ALEPH detector in 1997 at 181-184 GeV,
using $e^+e^-\rightarrow \gamma Z \rightarrow hadrons$ events.

\vspace{2cm}
\begin{center}
{ \em ALEPH contribution to 1998 Summer Conferences}\\
{\em Contact person: Eugeni Graug\'es (Eugeni.Grauges@cern.ch)}
\end{center}
\end{titlepage}
%-----------------------------------------------------------------------
\newpage
%-----------------------------------------------------------------------
\section{Introduction}
\label{intro}

One of the main goals of LEP2 is the direct measurement of the W mass
with an accuracy of 30-50 MeV, due to its implications in the validation 
of the Standard Model. At the same time, this requires a very 
precise determination of the LEP centre-of-mass energy, i.e., 
below 30 MeV. LEP1 was ensured to get this accuracy  due to 
the availability of beam transverse polarisation to apply the Resonant 
Depolarisation (RD) method for energy measurement.

At LEP2 this is not anymore the case and,
starting from a precise calibration of the RD method at low (LEP1) energies,
an extrapolation to higher energies is made, within which some precision 
may be lost. 

        Lately, an additional possibility has been considered, either as a 
cross-check on the previous method cited or as an alternative measurement 
if enough precision is proven .
 It consists in determining the LEP centre-of-mass energy 
by means of an offline analysis of the recorded events.

 Although at LEP2 the colliding energy is far above from the Z peak, 
the physics processes that take place are very sensitive to it.
 The presence of  the so-called ``Radiative Return Events'',
 where the initial colliding particles (electron or positron) radiate photons
 before interacting among themselves, is the responsible for such sensitivity.
 The Initial State
Radiation (ISR) is mostly collinear to the beam pipe thus, 
lost and not detected. Therefore, the effective 
centre-of-mass energy is reduced from the nominal one ($\sqrt{s}$),
 leading to a  ``scanning''  of the 
effective centre-of-mass energy ($\sqrt{s'}$), 
in which the Z resonance appears naturally.
 By analysing the two fermions final state processes it is possible 
to determine the
effective centre-of-mass energy and, the very precise 
knowledge of the Z mass and width 
 can be used as a strong constraint to perform a fit  to the line shape
in order to extract  the LEP
centre-of-mass energy.

In summer 1997 the nominal LEP centre-of-mass energy was raised up 
to 181, 182, 183 and 184 GeV, and an integrated luminosity of 56.812 pb$^{-1}$
was collected by ALEPH. This note describes a LEP centre-of-mass measurement by a line shape fit to the $q\bar{q}$ final state processes.

 The $q\bar{q}$ production cross-section at these energies
is large enough, more than 5000 $q\bar{q}$ events expected, to allow a measurement
of the LEP centre-of-mass with an accuracy of 1 per mil. 
The events are selected making use of very simple cuts which nevertheless lead
to a high purity with a reasonable signal efficiency. After possible 
ISR photons non-collinear with the beam are identified, the events are forced into
two jets. At this point, the effective centre-of-mass energy is extracted from the 
jet angles to improve the  resolution and to provide at the same time, a natural 
kinematical limit to the events. Finally, the experimental 
$x$ distributions \footnote{The $x$ 
variable  is equivalent to $s'$ since the relation $x=1-(s'/s)$. For technical
reasons $x$ distribution is  prefered than $s'$ distribution, to perform the fit.} 
are compared with Monte Carlo reweighted distributions to extract the value 
of the LEP centre-of-mass energy.

\section{Monte Carlo samples}
\label{mc} 
A sample of 100,000 $q\bar{q}$ events at a nominal CM energy of 182.675 GeV 
fully simulated using  KORALZ \cite{KRLZ}  were used to perform the data fit. Monte Carlo 
samples at 183 GeV with integrated luminosities very much larger that of the data, 
were simulated for all background reactions. The PYTHIA \cite{PYTH} 
generator was used to generate  WW, ZZ 
and Zee events. Two photon ($\gamma\gamma$) reaction into hadrons were 
simulated with the PHOT02 \cite{PHOT} generator.
Also for calibration of the method  purposes, $q\bar{q}$ samples of 350,000 at 183 GeV 
and 20,000 either at 182 as well as 184 GeV fully simulated with 
PYTHIA  generator were used.  

\section{Event selection and reconstruction algorithm}
\label{selection}
At $\sqrt{s}$ = 183 GeV the main backgrounds to the 
process $e^+e^- \rightarrow q \bar{q}$ are 
$e^+e^- \rightarrow  WW$, $e^+e^- \rightarrow  ZZ$,
$e^+e^- \rightarrow  \gamma\gamma$ and 
$e^+e^- \rightarrow  Zee$. In order to discriminate between signal and background, 
the following cuts were applied.
\begin{itemize}
\item{} Aleph  standard $q\bar{q}$ candidate selection: This selection 
                is performed using the ALEPH Energy Flow algorithm \cite{EFLW} associated to 
                a charged track preselection, which requires at least 5 TPC tracks 
                satisfying the following cuts: At least 4 TPC hits,  the 
                track  originated from within a cylinder with radius 2 cm and length 10 cm, 
                centered around the interaction point, $|cos\theta| <$  0.95. In addition the total 
                energy of all TPC tracks satisfying the previous cuts, should have more than 10\%
                of the nominal centre-of-mass energy.    
\item{} Total visible invariant mass greater 50 GeV
\item{} Number of charged tracks greater than 7
\end{itemize}

After that, ISR photon non collinear with the beam pipe and entering into 
the detector are identified as a thin electromagnetic jet.
 The remaining part of the events are then forced into two
jets using the DURHAM-P algorithm. To reconstruct the effective 
centre-of-mass energy $\sqrt{s'}$, it is assumed that the ISR photons are 
emitted along the beam pipe, boosting the centre-of-mass 
in that direction (except for the cases in which such ISR photons
 are indentified inside the detector, in which case  their direction is taken 
as the boost direction). Under that approximation the magnitude of the boost $\beta$
can be computed from the measured directions (polar angles $\theta_1$ 
and $\theta_2$) of the jets with respect to direction opposite to the photon.
 Assuming no Final State Radiation (FSR) 
and following the method  described in \cite{FEDE}, the effective
 centre-of-mass energy can be expressed as:  
\begin{eqnarray}
           s' = s~(1-x) 
\end{eqnarray}
where
\begin{eqnarray}
          x = \frac{2|\beta|}{1+|\beta|}
\end{eqnarray}
and 
\begin{eqnarray}
          |\beta| = \frac{|sin(\theta_1+\theta_2)|}{sin(\theta_1)+sin(\theta_2)}
\end{eqnarray}
   
Once $s'$, or equivalently $x$, is reconstructed  only the events included in the 
following $x$-window  are selected:
\begin{eqnarray}
 0.6 < x < 0.8775 
\end{eqnarray}
Using the MC samples generated for this study, the expected observable 
cross-section for each 
process  are summarised in Table~\ref{efftab}
\begin{table}[htb]
\begin{center}
\begin{tabular}{|c|c|}
\hline
Processes & $\sigma_{eff}$(pb) \\
\hline
$q\bar{q}$ &  59.1   \\
\hline
\hline
WW & 1.9 \\
\hline
ZZ & 0.8 \\
\hline
$\gamma\gamma$ & 1.1\\
\hline
Zee   & 1.5 \\
\hline
\hline
Purity & 91.8\% \\
\hline
\end{tabular}
\end{center}
\caption[]  
{\protect\footnotesize
Effective cross-section for various processes after selection cuts.}
\label{efftab}
\end{table}

%-----------------------------------------------------------------------

%-----------------------------------------------------------------------

\section{LEP Centre-of-Mass Energy measurement.}
\label{measurement}

The LEP centre-of-mass energy is determined from the 181, 182, 183 and 184 datasets
separately and then combined taking into account the relative luminosity of each dataset.
In each case, a Monte carlo  reweighting procedure developed earlier \cite{WMASS} is
employed to  find the value of E$_{cm}$ which best fits the reconstructed $x$ distribution. 
Selected Monte Carlo (KORALZ) signal events from the large sample at a reference 
energy of 182.675 GeV are reweighted using 
the differential production cross-section \footnote{Since we are reweighting MC events,
 both the reconstructed $x$ (as obtained in the data) and the true $x$ 
(from the effective centre-of-mass energy after ISR) are available event by event.}
\begin{eqnarray}
           \frac{{\it d\sigma}}{{\it dx}} (x_{true},E_{cm}) 
\end{eqnarray}
according to the parameter to be fitted, E$_{cm}$. Background Monte Carlo
 samples are included  in the fit, but they are 
not reweighted, i.e., their energy dependence is not taken into account. This may introduce
 an additional source of systematic error which is estimated in section~\ref{backsys}

The statistical error in E$_{cm}$ is computed from the single fits to the 
data distributions.
%, since the size of the data sample
% provide a negligible uncertainty in the statiscal error
%obtained from the fit.

In the present analysis of the 1997 data, the reweighting procedure is 
applied to the $x$ distribution. A log-likelihood fit is performed with fixed
bins of 0.01 over the $x$ range of 0.60-0.88.
% This matches exacly the x window  employed in the selection cut.  

\section{The results}
\label{results}
%--------------------------------------------
\begin{table}[htb]
\begin{center}
\begin{tabular}{|c|c|c|c|}
\hline
Nominal & Data   & Expected &  Integrated \\
dataset & events &  events  &  Luminosity (pb$^{-1}$)\\
\hline
181 &   17 &   11  & 0.166\\
\hline
182 &  267 &  253 & 3.924\\
\hline
183 & 3375 & 3269 & 50.795\\
\hline
184  & 123 &  124 & 1.927\\
\hline
\hline
All &  3782 &   3657 & 56.812          \\
\hline
\end{tabular}
\end{center}
\caption[]  
{\protect\footnotesize Number of events found in the data after the selection 
cuts at each nominal energy dataset, as well as the corresponding
number of expected events  and integrated luminosity   }
\label{data}
\end{table}
%------------------------------------------
%--------------------------------------------
\begin{table}[htb]
\begin{center}
\begin{tabular}{|c|c|c|c|}
\hline
Nominal & E$_{cm}$ & -$\Delta$E$_{cm}$ & +$\Delta$E$_{cm}$  \\
dataset & (GeV) & (GeV) & (GeV)  \\
\hline
181 & 183.486  & -2.839 &  3.472  \\
\hline
182 & 180.294  & -0.608 & 0.629  \\
\hline
183 & 182.541  & -0.219 &  0.221 \\
\hline
184  & 185.711 & -1.432 & 1.604 \\
\hline
\hline
Combined & 182.496 &  -0.206 & 0.210 \\
\hline
\end{tabular}
\end{center}
\caption[]  
{\protect\footnotesize Fitted LEP centre-of-mass energies with their respective 
statistical errors for each nominal dataset}
\label{energy}
\end{table}
%------------------------------------------------------
The number of events found in the data after the selection 
cuts at each nominal energy dataset, as well as the corresponding
number of expected events is summarized in table~\ref{data}. Also the 
integrated luminosity for each data taking period is provided.

The LEP centre-of-mass energy values and the statistical errors obtained from  single fits to
the different datasets is summarized in table~\ref{energy}, as well as the value coming  from
their combination taking into account the integrated luminosity of each dataset.

In table~\ref{nading}, the LEP centre-of-mass energy values  obtained for ALEPH by means of
 the standard LEP energy calibration procedure (i.e., extrapolation from RD) are summarized
for each nominal energy dataset. Also their combination, according to the respective luminosity 
of each data taking period, is quoted.
\begin{table}[htb]
\begin{center}
\begin{tabular}{|c|c|}
\hline
Nominal & E$_{cm}$   \\
dataset & (GeV)   \\
\hline
181 & 180.826 $\pm$ 0.050  \\
\hline
182 & 181.708 $\pm$ 0.050  \\
\hline
183 & 182.691 $\pm$ 0.050  \\
\hline
184  & 183.801 $\pm$ 0.050 \\
\hline
\hline
Combined & 182.652 $\pm$ 0.050 \\
\hline
\end{tabular}
\end{center}
\caption[]  
{\protect\footnotesize  LEP centre-of-mass energies obtained by means of the standard
LEP energy calibration procedure (i.e., extrapolation from RD) for each nominal dataset.}
\label{nading}
\end{table}

\begin{figure*}[htb!]
\vspace{-1.cm}
\begin{center}
\mbox{\psfig{file=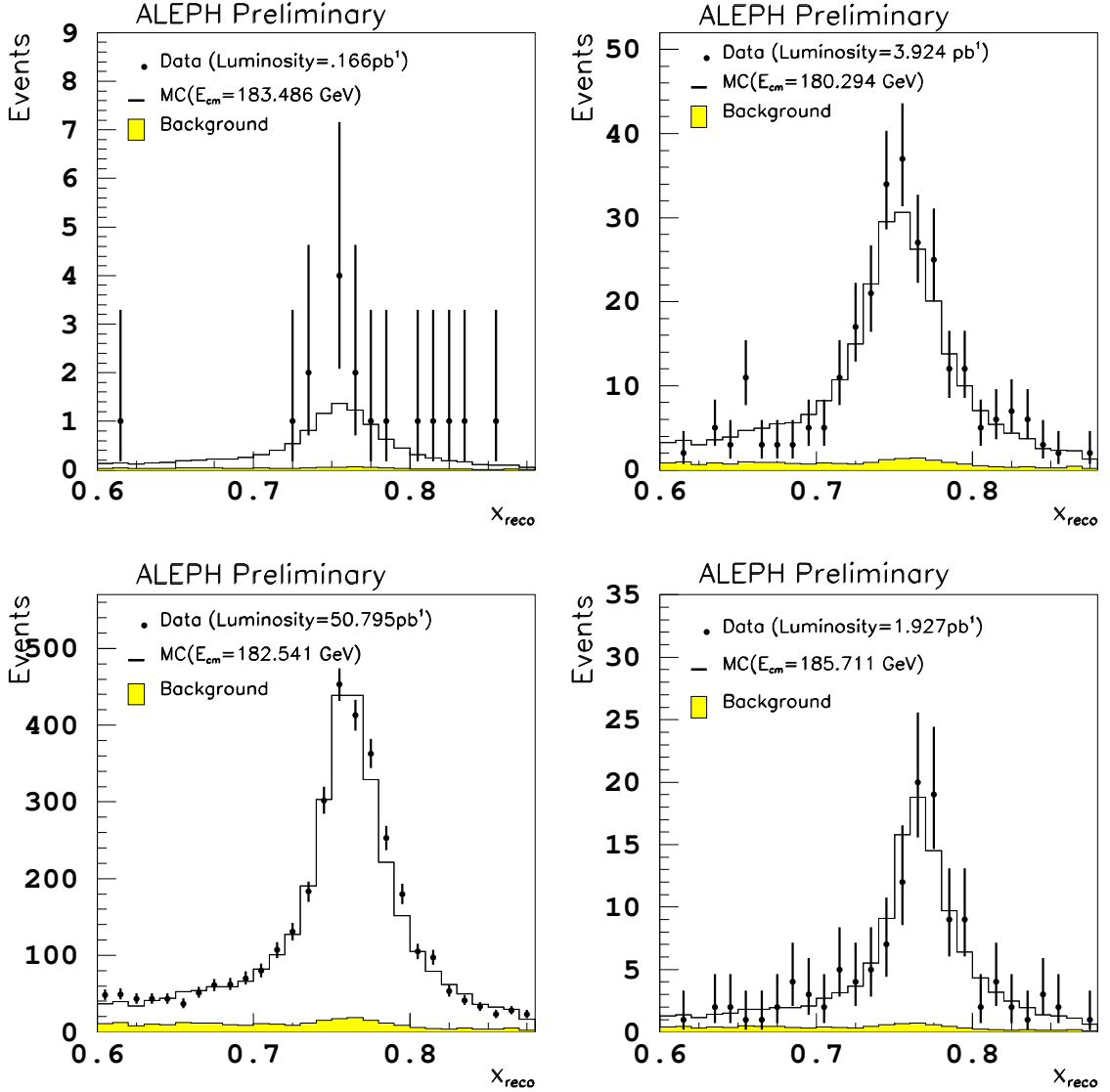,height=17cm}}
\end{center}
\vspace{-1.2cm}
\caption[]
{\protect\footnotesize
 Reconstructed x distribution ($x_{reco}$)  for data selected events (points with error
bars), background (shaded area) and signal+background Monte Carlo
for the best fit to the data (solid line histogram). The above 
figures correspond to the four 1997 nominal centre-of-mass energy datasets. 
The Monte Carlo reweighted prediction plotted makes use of the energy 
value which best fits the data, in each case}
\label{datafitfig}
\end{figure*}

Figure~\ref{datafitfig} shows the reconstructed $x$ ($x_{reco}$) distribution for the selected 
events in each nominal dataset, compared with the Monte Carlo reweighted prediction
using the LEP centre-of-mass energy wich best fits the data. The $\chi^2$ of the histograms range from 
36 up to 48 over 28 degrees of freedom. 

%--------------------------------------------------------------------

\section{Linearity of the reweighting technique}
\label{calibration}
\begin{figure*}[htb!]
\vspace{-1.cm}
\begin{center}
\mbox{\psfig{file=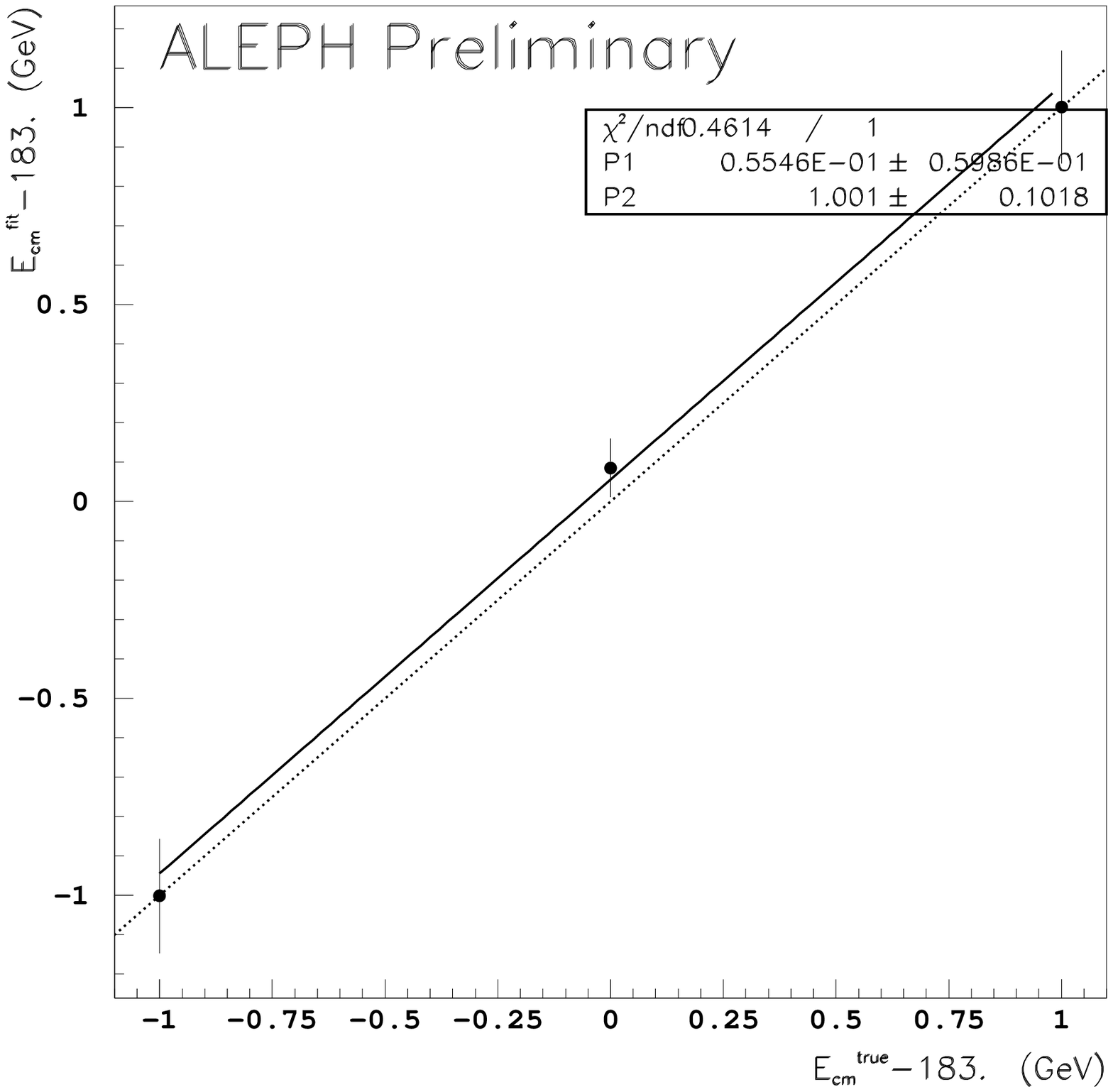,height=10cm}}
\end{center}
\vspace{-1.2cm}
\caption[]
{\protect\footnotesize Calibration curve of fitted against true energy. The slope 
is consistent with 1 meanwhile the offset is non significant.The full line 
corresponds to the result of the fit whereas the dotted line represents the ideal case
E$_{cm}$$^{fit}$=E$_{cm}$$^{true}$ }
\label{calibfig}
\end{figure*}

A critical test of the reweighting method is to ensure that the fitted 
energy agrees with the true  energy, when performing a fit to 
a Monte Carlo sample. The linearity of the fitted energy with the true input energy 
is studied for this $q\bar{q}$ channel using three independent Monte Carlo samples
with different input energies. The fitted values from these distributions have 
slope consistent with a  value of one,  1.001$\pm$0.102, and a non significant 
offset of 0.055$\pm$0.060 GeV is observed (see figure~\ref{calibfig}).

%---------------------------------------------------------------------
\section{Systematic uncertainties}

\subsection{Calorimeter calibrations}
        During the 1997 data taking the uncertainties in the global calibrations
of ECAL and HCAL energy were assessed to be $\pm$0.9\% and $\pm$2\% respectively. 
Adjusting the data in each direction for ECAL and HCAL by these uncertainties, the errors are
estimated by the shift obtained fitting the energy in each case. These 
are 39 MeV/c$^2$ and 55 MeV/c$^2$ respectively. The larger of the shifts 
 obtained from the fit is quoted in each case.

\subsection{Jet angle calibration}
Z peak data are used to map the response of the detector
to hadronic jets as a function of polar angles. The observed difference
in the directions of the two jets between data and Monte Carlo as a function of 
 $|cos\theta |$  is used to implement 
the corresponding correction to the data. The shift obtained from the 
fit of $E_{cm}$, which is 8 MeV,  is taken as the systematic error.

\subsection{MC statistics}
The finite number of Monte Carlo events used as a reference in the reweighting 
method contributes with a systematic uncertainty of 35 MeV.

\subsection{Initial State Radiation}

ISR is featured in KORALZ up to {\cal O}($\alpha^2L^2$), i.e. up to second 
order in the leading-log approximation, in the YFS \cite{YFS} style. 
The effect of the missing higher order terms  in the LEP centre-of -mass energy have been 
studied here by degrading KORALZ to {\cal O}($\alpha^1L^1$), and checking how 
large the pure {\cal O}($\alpha^2L^2$) correction is. 
This effect, about 18 MeV, is quoted as an upper limit to the systematic effect in the 
LEP centre-of-mass energy determination coming from these missing higher order terms.

\subsection{ISR-FSR interference}
There exists no Monte Carlo, for the time being, giving a good representation of Initial-Final State
interference for quarks final state. However an evaluation of the order of 
magnitude of the effect has been performed for muon final 
state using an {\cal O}($\alpha$)
simulation \cite{LEPEWGMEETING}. An upper limit of 50 MeV on E$_{cm}$ could be given.
Although the effect is presumably much smaller for quarks because of i) their 
charge ii) the interplay of gluon emission and fragmentation, these 50 MeV are taken 
as a preliminary systematic error. 
\subsection{Background contamination}
\label{backsys}
%For the $q\bar{q}$ events,
The expected background remaining after 
selection cuts is 8.2\%. Although small, the background contamination also depends
on the centre-of-mass energy. Since, in our case, the background is not reweighted
as the signal is, this is introducing an additional source of systematic error on 
 top of the intrinsic discrepancy between data and Monte Carlo at a 
given centre-of-mass energy. This effect can affect either the background shape, 
as well as its normalization. Therefore to estimate the systematic error coming 
from the background we assume uncertainties both in the total normalization 
and in the background shape. 
\subsubsection{Total normalization}
   Assuming  a very conservative $\pm$10\% uncertainty in the total background normalization, 
the reference MC sample has been adjusted in each direction. The larger of the shifts 
obtained is 32 MeV, and is quoted as the systematic error.
\subsubsection{Background shape}
To estimate the systematic error coming from the background shape 
very conservative uncertainties of $\pm$10\% are assumed for each background process independently.
Again, adjusting in each direction the normalization of each background process, the 
shifts are obtained in each case and summarized in table~\ref{backshape}. Also here, 
the larger of the uncertainties from the shifts applied are quoted in each case. 
The total systematic error due to the uncertainty of the background shape is estimated to be 37 MeV

\begin{table}[htb]
\begin{center}
\begin{tabular}{|c|c|}
\hline
Processes ($\pm$10\%) & Deviation(MeV) \\
\hline
WW & 35 \\
\hline
ZZ & 9 \\
\hline
$\gamma\gamma$ & 5\\
\hline
Zee   & 4 \\
\hline
\hline
Total & 37 \\
\hline
\end{tabular}
\end{center}
\caption[]  
{\protect\footnotesize Background shape uncertainty systematic error. Errors obtained assuming a
$\pm$10\% uncertainty in each background process independently.         }
\label{backshape}
\end{table}

\subsection{Summary of the systematic errors}

Table~\ref{syslist} lists all systematic errors considered.

\begin{table}[htb]
\begin{center}
\begin{tabular}{|c|c|}
\hline
Source & $\Delta$E$_{cm}$(MeV) \\
\hline
\hline
Calorimeter  calibrations & 67 \\
\hline
MC statistics      & 35 \\
\hline
ISR & 18\\
\hline
ISR-FSR Interference   & 50 \\
\hline
Background Contamination & 49 \\
\hline
Jet angle calibration   & 8 \\
\hline
\hline
Total &   105\\
\hline
\end{tabular}
\end{center}
\caption[]  
{\protect\footnotesize Summary of the systematic errors on the LEP centre-of-mass measurement }
\label{syslist}
\end{table}

\section{Summary and conclusions}
A Monte Carlo reweighting technique is used to measure the LEP centre-of-mass energy
from $Z\gamma$ events. It is based on the direct comparison of the  $x$ 
distributions ($x$=1- $s'/s$) reconstructed from the data
with those from Monte Carlo weighted events.
Selection cuts are applied to the data to study radiative return events in 
the $q\bar{q}$ final state channel, in order to get the maximum compromise between purity of the sample
and signal efficiency of the most sensitive events (whose $\sqrt{s'}$ is around the Z resonance).
        The resulting $x$ distribution for each nominal energy dataset collected by ALEPH is compared 
with  reweighted Monte Carlo distributions,
and the values of the LEP centre-of-mass energy are extracted in a binned log-likelihood method.
Combining all the measurements at different $\sqrt{s}$ taking into account their respective integrated luminosity, the average
LEP centre-of-mass energy at ALEPH for the high energy run of 1997, is measured to be:
\begin{eqnarray}
 E_{cm} = 182.496 \pm 0.208 ~(stat.) \pm 0.105 ~(syst)  ~GeV 
\end{eqnarray}
This result is consistent with the estimate from the LEP energy working
group~\cite{lepewg} weighted over the ALEPH luminosity events:
E$_{cm}$=182.652 $\pm$ 0.050 GeV.
With the expected increase in statistics at LEPII and with refined
experimental techniques akin to those used for the W mass
determination~\cite{alephmw183}, the technique described here should
provide an alternative measurement of the LEP center-of-mass energy,
with competitive errors. Already, if only this evaluation of the LEP 
centre-of-mass-energy were used in the
W mass determination, the systematic error on M$_W$ coming from the
beam energy scale would be
102 MeV, which is smaller than the experimental  error of 139 MeV
on M$_W$ from the same data sample~\cite{alephmw183}.
 
%\section{Acknowledgements}

%
%--------------------------------------------------------------------------
% Bibliography
%--------------------------------------------------------------------------

%
%%%%%%%%%%%%%%%%%%%%%%%%%%%%%%%%%%%%%%%%%%%%%%%%%%%%%%%%%%%%%%%%%%%%%%%%%%%%%
% FIGURES
%%%%%%%%%%%%%%%%%%%%%%%%%%%%%%%%%%%%%%%%%%%%%%%%%%%%%%%%%%%%%%%%%%%%%%%%%%%%%
%%%%%%%%%%%%%%%%%%%%%%%%%%%%%%%%%%%%%%%%%%%%%%%%%%%%%%%%%%%%%%%%%%%%%%%%%%%%
%\begin{figure}[p]
%\begin{center}
%\mbox{
%\epsfig{file=file.eps,width=12cm}
%}
%\end{center}
%\caption[]
%{\protect\footnotesize
% xxxx.}
%\label{}
%\end{figure}
%%%%%%%%%%%%%%%%%%%%%%%%%%%%%%%%%%%%%%%%%%%%%%%%%%%%%%%%%%%%%%%%%%%%%%%%%%%%

%
\end{document}